\documentclass[preprint,authoryear,12pt]{elsarticle}

\usepackage{amssymb}

\usepackage{url}  
\usepackage{ifthen}  
\usepackage{multicol}   
\usepackage[utf8]{inputenc} 
\usepackage{graphicx}
\usepackage{amssymb}
\usepackage{eucal}
\usepackage[english]{babel}
\usepackage{microtype}
\usepackage{amsmath}    
\usepackage{amsfonts}
\usepackage{mathrsfs}
\usepackage{afterpage}
\usepackage{tabularx,colortbl}
\usepackage{tikz}
\usetikzlibrary{automata}
\usepackage[caption=false]{subfig}
\usepackage{pgf}
\usepackage{latexsym}
\usepackage{stmaryrd}
\usepackage{mathrsfs}
\usepackage{algorithm}
\usepackage{algorithmicx}
\usepackage{algpseudocode}
\usepackage{booktabs}

\journal{Pattern Recognition Letters}

\begin{document}


\begin{frontmatter}



\title{Towards an Integrated Approach to Crowd Analysis and Crowd Synthesis: a Case Study and First Results}


\author{Stefania Bandini\textsuperscript{1}, Andrea Gorrini\textsuperscript{2}, Giuseppe Vizzari\textsuperscript{1}\corref{corr}}

\cortext[corr]{Corresponding author}

\ead{bandini@disco.unimib.it, a.gorrini@campus.unimib.it, vizzari@disco.unimib.it}

\address{\textsuperscript{1}Complex Systems and Artificial Intelligence Research centre,\\
Universit\`a degli Studi di Milano--Bicocca, Milano, Italy\\

\textsuperscript{2}Information Society Ph.D. Program, Department of Sociology and Social Research,  Universit\`a degli Studi di Milano--Bicocca, Milano, Italy}

\begin{abstract}
Studies related to \emph{crowds} of pedestrians, both those of theoretical nature and application oriented ones, have generally focused on either the \emph{analysis} or the \emph{synthesis} of the phenomena related to the interplay between individual pedestrians, each characterised by goals, preferences and potentially relevant relationships with others, and the environment in which they are situated. The cases in which these activities have been systematically integrated for a mutual benefit are still very few compared to the corpus of crowd related literature. This paper presents a case study of an integrated approach to the definition of an innovative model for pedestrian and crowd simulation (on the side of synthesis) that was actually motivated and supported by the analyses of empirical data acquired from both experimental settings and observations in real world scenarios. In particular, we will introduce a model for the adaptive behaviour of pedestrians that are also members of groups, that strive to maintain their cohesion even in difficult (e.g. high density) situations. The paper will show how the synthesis phase also provided inputs to the analysis of empirical data, in a virtuous circle.

\end{abstract}

\begin{keyword}

crowd analysis \sep crowd synthesis \sep agent--based modelling and simulation \sep groups


\MSC[2010]{68T42,68U20}

\end{keyword}

\end{frontmatter}


\section{Introduction}\label{sec:intro}

The modelling and simulation of pedestrians and crowds is a consolidated and successful application of research results in the more general area of computer simulation of complex systems. It is an intrinsically interdisciplinary effort, with relevant contributions from disciplines ranging from physics and applied mathematics to computer science, often influenced by (and sometimes in collaboration with) anthropological, psychological, sociological studies and the humanities in general. The level of maturity of these approaches was sufficient to lead to the design and development of commercial software packages, offering useful and advanced functionalities to the end user (e.g. CAD integration, CAD-like functionalities, advanced visualisation and analysis tools) in addition to a simulation engine\footnote{see \url{http://www.evacmod.net/?q=node/5} for a large although not necessarily complete list of pedestrian simulation models and tools. The list comprises more than 60 models, of commercial and academic nature, general purpose or specifically designed for certain situations and scenarios, maintained or discontinued.}. Nonetheless, as testified by a recent survey of the field by~\cite{DBLP:reference/complexity/SchadschneiderKKKRS09} and by a report commissioned by the Cabinet Office by~\cite{UnderstandingCrowd}, there is still much room for innovations in models improving their performances both in terms of \emph{effectiveness} in modelling pedestrians and crowd phenomena, in terms of \emph{expressiveness} of the models (i.e. simplifying the modelling activity or introducing the possibility of representing phenomena that were still not considered by existing approaches), and in terms of \emph{efficiency} of the simulation tools. In addition to the above directions, we want to emphasise the fact that one of the sometimes overlooked aspects of a proper simulation project is related to the \emph{calibration} and \emph{validation} of the results of tools related to the \emph{synthesis} of the pedestrians and crowd behaviour in the considered scenario. These phases are essentially related to the availability of proper empirical data about or, at least, relevant to, the considered scenario ranging from the pedestrian demand (i.e. an origin--destination matrix), preferences among different alternative movement choices (e.g. percentage of persons employing stairs, escalators and elevators in a multiple--level scenario), but also the average waiting times at service points (i.e. queues), the average time required to cover certain paths, the spatial distribution of pedestrians in specific environmental conditions that is required to evaluate the so--called ``level of service'' associated to portions of the environment as defined by~\cite{fruin1971pedestrian}. These data are results of activities of \emph{analysis}, some of which can be fruitfully automated given, on one hand, the wide diffusion of cameras employed for video surveillance of public areas and, on the other, considering the level of maturity of video processing and analysis techniques.

An \emph{integrated approach} to pedestrians and crowd studies encompasses both the application of analysis and synthesis techniques that, in a virtuous circle, can mutually benefit one from the other, to effectively (i) identify, (ii) face and (iii) provide innovative solutions to challenges towards the improvement of the understanding of crowding phenomena. This kind of interdisciplinary study, in addition to computer science, often employs or directly involves research results in the area of social sciences in general, can be found in the literature. For instance,~\cite{DBLP:journals/tvcg/PatilBCLM11} show how computational fields guiding simulated pedestrians movement in a simulated environment can be automatically derived by video footages of actual people moving in the same space. \cite{DBLP:journals/cacm/MooreAMS11}, instead, employ an hydrodynamic model, that (to a certain extent) can represent the flow of pedestrians in mass movement situations, to improve the characterisation of pedestrian flows by means of automatic video analysis. From the perspective of offering a useful service to crowd managers~\cite{DBLP:journals/sj/GeorgoudasSA11} describe an anticipative system integrating computer vision techniques and pedestrian simulation able to suggest crowd management solutions (e.g. guidance signals) to avoid congestion situations in evacuation processes. Finally, in~\cite{DBLP:conf/hbu/RaghavendraBCM11} the authors propose the employment of the \emph{social force model} by~\cite{HelbingSocialForce}, probably the most successful example of crowd synthesis model, to support the detection of abnormal crowd behaviour in video sequences.

An example of identification of a still not considered phenomenology is related to a work by~\cite{TheraulazGroup} in which the authors have defined an extension to the social force model by that considers the presence of groups in the simulated population: the motivations and some modelling choices (i.e. the limited size of considered groups and their spatial arrangement) are based on actual observations and analyses. A related effort, carried out instead by a research group trying to improve crowd analysis techniques, is described in~\cite{DBLP:conf/iccvw/Leal-TaixePR11}: in this case, the social force model acts as a sort of predictor block in an automated video analysis pipeline, improving the tracking in case of groups within the observed flow. Finally,~\cite{Group-Video-Analysis-Schultz} also focus on groups, as central element of an observation and analysis that also considers psychometric factors.

It is important to emphasise that anthropological considerations about human behaviour in crowded environments, such as the analysis of spatial social interaction among people, are growingly considered as crucial both in the computerised analysis of crowds as pointed out by~\cite{CrowdAnalysisComputerVision} and in the synthesis of believable pedestrian and crowd behaviour, such as in~\cite{DBLP:conf/icaisc/Was10} and in~\cite{VizAamas2011}. In particular, \textit{proxemics} has a prominent role both in the modelling and analysis of pedestrians and crowd behaviour. The term was first introduced by~\cite{Hall-HiddenDimension} with respect to the study of a type of non--verbal behaviour related to the dynamic regulation of interpersonal distance between people as they interact. 

Within this context the aim of this paper is to provide a comprehensive framework comprising both the synthesis and analysis of pedestrians and crowd behaviour: in this schema we suggest both ways in which the results of the analysis can provide fruitful inputs to modellers and, on the other hand, how results of the modelling and simulation activities can contribute to the (automated) interpretation of raw empirical data. The framework will be described in the following section, while an example of unfolding of these conceptual and experimental pathways will be described through the introduction of an adaptive model for group cohesion (Section~\ref{sec:model}) that was motivated and that will effectively be calibrated and validated by means of analyses on observed crowd behaviour (Section~\ref{sec:observation}). Conclusions and future works will end the paper.

\section{An Integrated Framework for Crowd Analysis and Synthesis}\label{sec:framework}

A comprehensive framework trying to put together different aspects and aims of pedestrians and crowd dynamics research has been defined in~\cite{CrowdAnalysisComputerVision}. The central element of this schema is the mutually influencing (and possibly motivating) relationship between the above mentioned efforts aimed at synthesising crowd behaviour and other approaches that are instead aimed at analysing field data about pedestrians and crowds in order to characterise these phenomena in different ways. It must be noted, in fact, that some approaches have the goal of producing aggregate level quantities (e.g. people counting, density estimation), while others are aimed at producing finer-grained results (i.e. tracking people in scenes) and some are even aimed at identifying some specific behaviour in the scene (e.g. main directions, velocities, unusual events). The different approaches adopt different techniques, some performing a \emph{pixel--level analysis}, others considering larger patches of the image, i.e. \emph{texture--level analysis}; other techniques require instead the detection of proper objects in the scene, a real \emph{object--level analysis}.

From the perspective of the requirements for the synthesis of quantitatively realistic pedestrian and crowd behaviour, it must be stressed that both aggregate level quantities and granular data are of general interest: a very important way to characterise a simulated scenario is represented by the so called fundamental diagram by~\cite{DBLP:reference/complexity/SchadschneiderKKKRS09}, that represents the relationship in a given section of an environment between the flow of pedestrians and their density. Qualitatively, a good model should be able to reproduce an empirically observed phenomenon characterised by the growth of the flow until a certain density value (also said \emph{critical density}) is reached; then the flow should decrease. However, every specific situation is characterised by a different shape of this curve, the position of critical density point and the maximum flow level; therefore even relatively ``basic'' counting and density estimation techniques can provide useful information in case of observations in real world scenarios. Density estimation approaches can also help in evaluating qualitatively the patterns of space utilisation generated by simulation models against real data. Tracking techniques instead can be adopted to support the estimation of traveling times (and length of the followed path) by pedestrians. Crowd behaviour understanding techniques can help in determining main directions and the related velocities.

\begin{figure}[t]
\centering
\includegraphics[width=.83\textwidth]{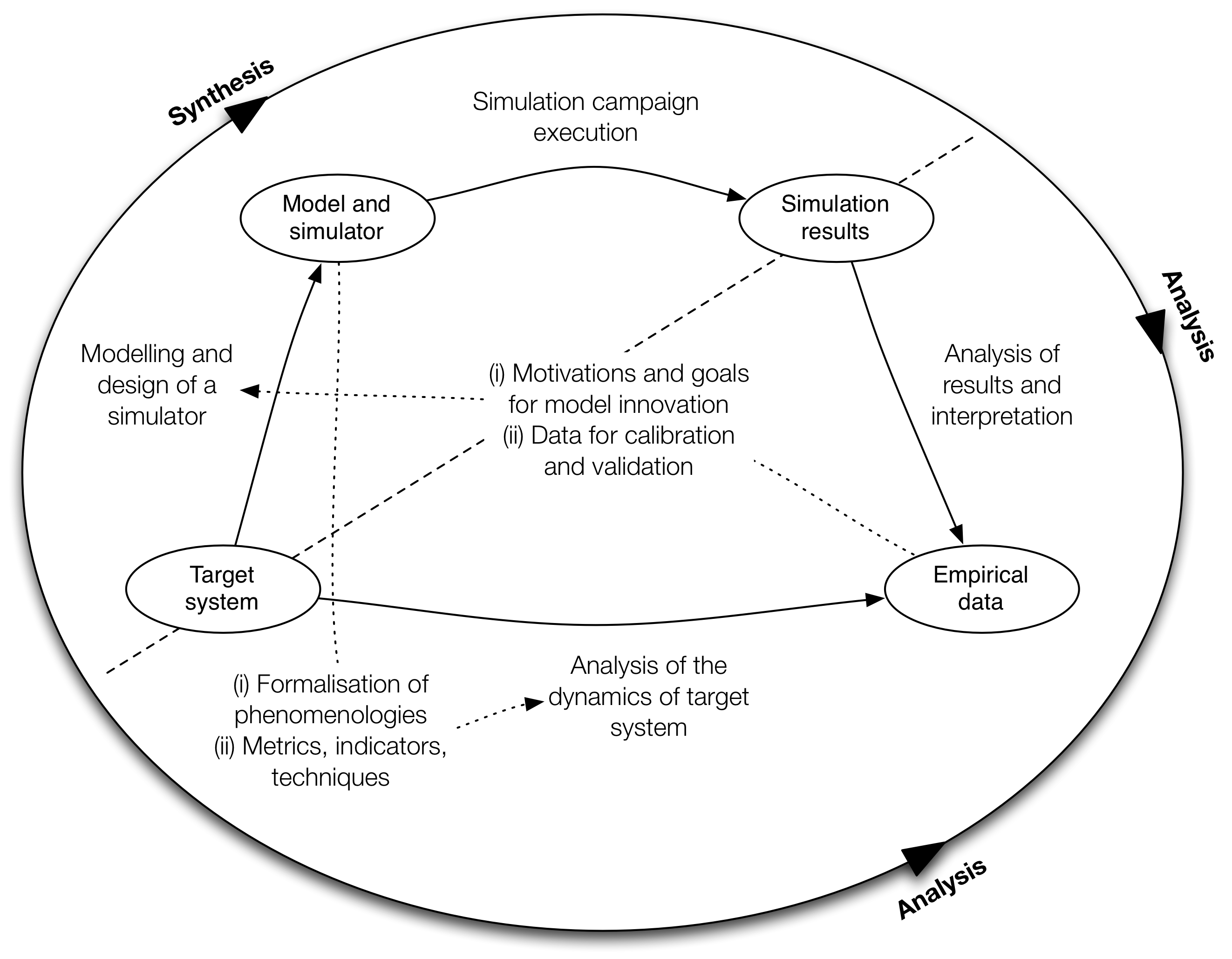}
\caption{A schema putting together activities of modelling and execution of simulation campaigns (synthesis) and activities related to the interpretation of simulation results, analysis of empirical data observed in reality and their comparison for sake of validation (analysis).}
\label{fig:framework}
\end{figure} 

To complete the initial schema, suggesting that analysis and synthesis should mutually benefit one from the other, we propose here the extension of a different kind of diagram, that can be found in~\cite{GilbertSimSocial} or more recently in~\cite{BanManVizJASSS}, used to discuss the form of inference that can be carried out by using a model as a method of study. Starting from portion of the reality that we will call \emph{target system} the process of synthesis leads to the definition of a model and its implementation in the form of a simulator. The latter can then be employed (i.e. executed with specific inputs and parameters) to carry out a simulation campaign leading to a set of results. The processes of analysis involve, on one hand, raw data that can be acquired through direct observations and controlled experiments. On the other hand, also simulation results require a process of interpretation in order to be comparable with observed empirical data. When this cycle produces simulation results that, once interpreted and analysed, actually match the empirical data acquired on the field, the defined model can be employed for sake of explanation and prediction. 

Of course it is not immediate to define a model that generates simulation results matching  empirical data, especially since models of complex systems are generally characterised by a number of parameters even when the modeller tries to keep the model as simple and elegant as possible. It is this need of actually \emph{calibrating} model parameters for achieving a model \emph{validation}, as described by~\cite{KluglSAC08}, that actually introduces the first type of synergetic collaboration between analysis and synthesis: the analysis of raw data about the simulated phenomenon leads to the possibility of identifying values or at least intervals for model parameters. This is not actually the only case of influence of the analysis on the definition of models: in fact, it is the observation of the system that leads to the identification of phenomenologies that are not currently represented and managed by a model and that can represent the motivations and goals for model innovation.

The potential outcomes of the modelling and simulation phases that can have an influence on the analysis activities are related to two categories of contributions: on one hand, the need to create a mechanism for the generation of an observed phenomenon leads to its formalisation, that could be instrumental also in the creation of additional mechanisms for its automated analysis. On the other hand, even long before reaching the necessary level of maturity of the simulator, the modeller and developer of a simulation system actually needs to define and develop metrics, indicators and techniques to evaluate the outcomes of the modelling phases. These by-products of the synthesis activity can also represent a starting point for the actual development of automated analysis approaches.

The following sections actually represent an unfolding of this kind of conceptual and experimental process. In particular, Section~\ref{sec:model} introduces a model for pedestrian and group behaviour representing the evolution of a first approach that was started to face issues raised by an unstructured and non-systematic observation of crowd patterns and movements in a real world scenario\footnote{In particular, the crowd management procedures adopted in the Arafat I station in Saudi Arabia, that adopt the notion of group as a way to organize the flow of pilgrims towards the station.} described by~\cite{VizManATT2012}. In this line of research, a model encompassing groups as a fundamental element influencing the overall system dynamics was designed and implemented and a metric for the evaluation of group cohesion (and, therefore, also its dispersion) was defined. This metric led to the understanding that the first version of the model was unable to preserve the cohesion of groups and therefore it was instrumental in the realisation of a new model described in~\cite{CASM2013} that endows pedestrians that are also members of specific types of group with an adaptive mechanism to preserve their cohesion even in difficult situations (e.g. presence of obstacles or high density environments). This research effort also led to the identification of new observations and analyses to back-up simulation results with empirical data; the dispersion metric was also the starting point for the formal definition of measurements to be executed in the analysis of the acquired raw data. These activities will be described in Section~\ref{sec:observation}.


\section{An Adaptive Model for Group Cohesion}\label{sec:model}

This section introduces a model representing pedestrian behaviour in an environment, considering the impact of the presence of simple and structured groups. The model is characterised by a discrete representation of the environment and time evolution, and it is based on the floor-field mechanism of existing CA approaches. However, the pedestrian behaviour is so articulated, comprising an adaptive mechanism for the preservation of group cohesion, to the point that the model is more properly classified as agent-based. The different elements of the model will now be introduced, then some results of its application to sample simulation scenarios will be given to show the model capabilities and the requirements in terms of empirical data to complete the calibration of the group cohesion adaptive mechanism. A more detailed formal introduction of the model and additional simulation results can be found in~\cite{CASM2013}.

\subsection{Representation of the Environment}\label{sec:environment}

The physical environment is represented in terms of a discrete grid of square cells: $Env = {c_0, c_1, c_2, c_3, ...}$ where $\forall c_i: c_i \in Cell$. The size of every cell is $40cm \times 40cm$ according to standard measure used in the literature and derived from empirical observation and experimental procedure shown in \cite{weidmann} and \cite{fruin1971pedestrian}. Every cell has a row and a column index, which indicates its position in the grid: $Row(c_i), Col(c_i): Cell \to \mathbb{N}$. Consequently, a cell is also identified by its row and column on the grid, with the following notation: $Env_{j,k} = c : (c \in Env) \; \land \; (Row(c) = j) \; \land \; (Col(c) = k)$. Every cell is linked to other cells, that are considered its neighbours according to the Moore neighbourhood. Cells are annotated and virtual grids are superimposed on the base environmental representation to endow the environment with the capability to host pedestrian agents and support their perception and action.

\subsubsection{Definition of Spatial Markers} 

Markers are sets of cells that play particular roles in the simulation. Three kinds of marker are defined in the model: (i) \emph{start areas}, places (sets of cells) were pedestrians are generated; they are characterised by information for pedestrian generation both related to the type of pedestrians and to the frequency of generation. In particular, a start area can generate different kinds of pedestrians according to two approaches: (a) \emph{frequency-based generation}, in which pedestrians are generated during all the simulation according to a frequency distribution; (b) \emph{en-bloc generation}, in which a set of pedestrians is generated at once in the start area when the simulation starts; (ii) \emph{destination areas}, final places where pedestrians might want to go; (iii) \emph{obstacles}, non-walkable cells defining obstacles and non-accessible areas.

\subsubsection{Definition of Floor Fields}

Adopting the approach of the \emph{floor field} model by~\cite{nishinari2004extended}, the environment of the basic model is composed also of a set of superimposed virtual grids, structurally identical to the environment grid, that contains different floor fields that influence pedestrian behaviour.

The goal of these grids is to support long range interactions by representing the state of the environment (namely, the presence of pedestrians and their capability to be perceived from nearby cells). In this way, a local perception for pedestrians actually simply consists in gathering the necessary information in the relevant cells of the floor field grids. 

Floor fields are either \emph{static} (created at the beginning and not changing during the simulation) or \emph{dynamic} (changing during the simulation). Three floor fields are considered in the model:
\begin{itemize}
\item the \textit{path field} assigned to each destination area: this field indicates for every cell the distance from the destination, and it acts thus as a gradient, a sort of potential field that drives pedestrians towards it (static floor field);
\item  the \textit{obstacles field}, that indicates for every cell the fact that an obstacle or a wall is within a limited distance, being maximum in cells adjacent to an obstacle and decreasing with the distance (static floor field);
\item the \textit{density field} that indicates for each cell the pedestrian density in the surroundings at the current time-step, analogous to the concept of \emph{cumulative mean density} (CMD) indicating the density experienced by pedestrians in a portion of the environment, first introduced by~\cite{KeithStillPhD} and elaborated for discrete environments by~\cite{StepsVsLegion} (dynamic floor field).
\end{itemize}

\subsection{Simulation Time and Update Strategy}

Simulation time is modelled in a discrete way by dividing time into steps of equal duration: we assume that a pedestrian moves (at most, since it is possible to decide to stand still) 1 cell per time step. The average velocity of a pedestrian, which can be estimated in real observations or experiments as performed by~\cite{fruin1971pedestrian} in about $1.2$ $ms^{-1}$, will thus determine the duration of the each time step: considering that the size of the cell is $40\textit{  cm} \times 40 \textit{ cm}$, the duration of each time step is $0.33 s$.


When running a CA-based pedestrian model, three update strategies are possible~\cite{KlupfelPhD}:
\begin{itemize}
	\item \emph{parallel} update, in which cells are updated all together;
	\item \emph{sequential} update, in which cells are updated one after the other, always in the same order;
	\item \emph{shuffled sequential update}, in which cells are updated one after the other, but with a different order every time.
\end{itemize}

The second and third update strategies lead to the definition of asynchronous CA models (see~\cite{DBLP:journals/nc/BandiniBV12} for a more thorough discussion on types of asynchronicity in CA models). In crowd simulation CA models, parallel update is generally preferred, as mentioned by~\cite{DBLP:reference/complexity/SchadschneiderKKKRS09}, even if this strategy can lead to conflicts that must be solved. Nonetheless, we currently adopted a shuffled sequential update scheme for a first evaluation of the group cohesion mechanism without adding additional mechanisms and parameters to be calibrated for the management of potential conflicts between agents' movements.

\subsection{Groups}

We focus on two types of group: simple and structured. Simple or informal groups are generally made up of friends or family members and they are characterised by a high cohesion level, moving all together due to shared goals and to a continuous mechanism of adaptation of the chosen paths to try to preserve the possibility to communicate, as discussed by~\cite{CostaGroupDistances}. Structured groups, instead, are more complex entities, usually larger than simple groups (more than 4 individuals) and they can be considered as being composed of sub-groups that can be, in turn, either simple or structured. Structured groups are often artificially defined with the goal of organising and managing a mass movement (or some kind of other operation) of a set of pedestrians. 

Groups can be formally described as:
\[
Group_j = \left\langle  Id, [Group_1, \ldots ,Group_m], [Ped_1, \ldots, Ped_n] \right\rangle 
\]

Structured groups include at least one subgroup, while simple groups only comprise individual pedestrians. We will refer to the group an agent $a$ \emph{directly} belongs to as $G_a$, that is also the smallest group he belongs to; the \emph{largest} group an agent $a$ belongs to will instead be referred to as $\bar{G_a}$. It must be noted that $\bar{G_a} = G_a$ only when the agent $a$ is member of a simple group that is not included in any structured group. These sets are relevant for the computation of influences among members of groups of different type (simple or structured).

\subsection{Pedestrians}
\label{sec:pedestrian}

In this model, a pedestrian is defined as an utility-based agent with state. Functions are defined for utility calculation and action choice, and rules are defined for state-change.
Pedestrians are characterised as:

\[
Pedestrian : \langle Id, GroupId, State, Actions, Destination \rangle
\]

where: (i) $Id \in \mathbb{N}$ is the agent identification number and $GroupId \in \mathbb{N}$ is the identification number of the group to which the pedestrian directly belongs to (for pedestrians that are not member of any group this value is null); (ii) $State$ is defined as: $State=\langle Position, PrevDirection \rangle$ where $Position$ indicates the current cell in which the agent is located, and $PrevDirection$ is the direction followed in the last movement; $Actions$ is the set of possible actions that the agent can perform, essentially movements in one of the eight neighbouring cells (indicated as cardinal points), plus the action of remaining in the same cell (indicated by an `$X$'): $Actions={N, S, W, E, NE, SE, NW, SW, X}$; $Destination$ is the goal of the agent in terms of destination area. This term identifies the current destination of the pedestrian: in particular, every destination is associated to a particular spatial marker. Consequently, $Destination$ is used to identify which path field is relevant for the agent: $currentPathField = PathField(Destination)$ where $PathField$ is the precise path field associated to $Destination$ and $currentPathField$ is the path field relevant for the agent.

All these elements take part in the mechanism that manages the movement of pedestrians: agents associate a desirability value, a utility to every movement in the cell neighbourhood, according to a set of factors that concur in the overall decision. 

\subsubsection{Mechanism of Action Evaluation}
\label{sec:movement}

In Algorithm \ref{alg1} the agent life-cycle during all the simulation time is proposed:  every time step, every pedestrian perceives the values of path field, obstacle field and density field for all the cells that are in its neighbourhood. On the basis of these values and according to different factors, the agent evaluates the different cells around him, associating an utility value to every cell and selects the action for moving into a specific cell.

\begin{algorithm}
\caption{Agent life-cycle}
\label{alg1}
\begin{algorithmic}
\ForAll{$timestep \in SimulationTime$}
\ForAll{$p \in Pedestrian$}
\State $Utility[]$
\ForAll{c $\in neighbours(Position)$} 
\State $\textit{pf} \gets Val(PathF,c)$
\State $\textit{of} \gets Val(ObsF,c)$
\State $df \gets Val(DensF,c)$
\State $\textit{Utility}[c] \gets Evaluation(pf,of,df)$
\EndFor
\State $a = \textit{Choice}(\textit{Utility}[])$
\State $Move(a)$
\EndFor
\EndFor
\end{algorithmic}
\end{algorithm}

As previously suggested, the action selection strategy starts gathering the value of floor fields in cells included in the neighbourhood of agent's current position. The obtained values will be used in the evaluation of the movement towards the related cell.

After acquiring the perceived information from the environment, the agent elaborates a desirability value for each of the admissible actions (movements), according to several factors, that will be described later on. Given the list of possible actions and associated utilities, an action is chosen with a probability proportional to its utility. In particular, the probability for an agent $a$ of choosing an action associated to the movement towards a cell $c$ is given by the exponential of the utility, normalised on all the possible actions the pedestrian can take in the current turn:

\[
p_a(c) = N \cdot e^{U_a(c)}
\]

where $N$ is the normalisation factor and $c$ is the currently considered destination cell. The utility function $U_a(c)$ of a destination cell $c$ which corresponds to an action/direction for agent $a$, that takes the form of a weighted sum of components associated to the behavioural specification of an agent:

$$U_a(c)=\frac{\kappa_g G(c)+\kappa_{ob} Ob(c)+\kappa_s S(c)+\kappa_{d} D(c)+\kappa_{ov} Ov(c)+\kappa_{c} C_a(c)+\kappa_{i} I_a(c)}{d}$$

where $d$ is the distance of the new cell from the current position, that is 1 for cells in the Von Neumann neighbourhood (vertically and horizontally neighbour cells) and $\sqrt{2}$ for diagonal cells: the factor is introduced to penalise the diagonal movements. 

The different components of the utility value for a given movement consider the following factors:
\begin{itemize}
\item the desire to move towards a \emph{goal}, a destination in the environment, represented by the $G(c)$ function;
\item the tendency to stay at a distance from the \emph{obstacles} (e.g. walls, columns), that are perceived as repulsive, represented by the $Ob(c)$ function;
\item the desire to stay at a distance from other individuals, especially those that are not members of the same simple group, an effect of proxemic \emph{separation}, represented by the $S(c)$ function;
\item a \emph{direction inertia} factor, increasing the desirability of performing straight forms of movement, represented by the $D(c)$ function;
\item the penalisation of those movements that cause an \emph{overlapping} event, the temporary sharing of the same cell by two distinct pedestrians, represented by the $Ov(c)$ function;
\item two contributions related to the tendency to preserve \emph{group cohesion}, respectively devoted to simple and structured groups, respectively represented by the $C_a(c)$ and the $I_a(c)$ functions.
\end{itemize}

Note that $\kappa_g, \kappa_{ob}, \kappa_s, \kappa_{d}, \kappa_{ov}, \kappa_{c}, \kappa_{i} \in [0, 100]$: the use of these parameters, in addition to allowing the calibration and the fine tuning of the model, also supports the possibility of describing and managing different types of pedestrian, or even different states of the same pedestrian in different moments of a single simulated scenario.

\subsubsection{Adaptation Mechanism for Group Cohesion Preservation}

While the above elements are sufficient to generate a pedestrian model that considers the presence of groups, even structured ones, the introduced mechanisms are not sufficient to preserve the cohesion of simple groups, as discussed in a previous work adopting a very similar approach (see~\cite{DBLP:conf/aiia/BandiniRVS11}). This is mainly due to the fact that in certain situations pedestrians adapt their behaviour in a more significant way than what is supported by simple and relatively small modifications of the perceived utility of a certain movement. In certain situations pedestrians perform an adaptation that appears in a much more decisive way a \emph{decision}: they can suddenly seem to temporarily loose interest in what was previously considered a destination to reach and they instead focus on moving closer to (or at least do not move farther from) members of their group, generally whenever they perceive that the distance from them has become excessive. In the following, we will discuss a metric of group dispersion that we adopted to quantify this perceived distance and then we will show how it can be used to adapt the weights of the different components of the movement utility computation to preserve group cohesion. 

\noindent \textbf{Group Dispersion Metrics} -- Intuitively, the dispersion of a group can be seen as the degree of spatial distribution of their members. In the area of  pedestrian modelling and simulation, the estimation of different metrics for group dispersion has been discussed in \cite{DBLP:conf/aiia/BandiniRVS11} in which different approaches are compared to evaluate the dispersion of groups through their  movement in the environment. In particular, two different approaches are compared here: (i) dispersion as occupied area and (ii) dispersion as distance from the centroid of the group. This topic was also considered in the context of computer vision algorithms such as in~\cite{Group-Video-Analysis-Schultz}, in which however essentially only \emph{line abreast} patterns were analysed. Therefore we will focus on the former approach.

Formally, the above introduced formulas of group dispersion for each approach are defined as follows:


\[
Disp(Group) = \displaystyle \frac{Area(Group)}{Size(Group)} \ \ \ \ \text{(Area method)}
\]
\[
Disp(Group) = \displaystyle \frac{\sum_{i=1}^{Size(Group)}distance(centroid,a_i)}{Size(Group)} \ \ \ \ \text{(Centroid method)}
\]

with $Area(Group)$ as the area occupied by the group, $Size(Group)$ as the number of its members, $centroid$ as its centroid.

The second metric appears much more straightforward when a continuous representation of the environment is possible or at least not in contrast with the adopted modelling approach: in the case of a discrete and relatively coarse discretisation its results are not particularly different from the first metric, but they are sometimes counterintuitive especially when describing particular group shapes (e.g. river-like lanes that are often present in high density situations).

The first metric defines the dispersion of the group as the portion of space occupied by the group with respect to the size of the group: the first step of the procedure of computation for this metric builds a convex polygon with the minimum number of edges that contain all the vertices (representing the position of a pedestrian); the second step computes the area of this polygon. The dispersion value is calculated as the relationship between the polygon area and the size of the group.


\noindent \textbf{Utility Parameters Adaptation}\label{sec:balance} -- The adopted approach is characterised by a trade-off process between the goal attraction value and the intra/inter cohesion value in the utility computation: in the situation in which the spatial dispersion value is low, the cohesion behaviour has to influence pedestrian's overall behaviour less than the goal attraction. On the contrary, if the level of dispersion of a group is high, the cohesion component for the members must become more important than the goal attraction. An adaptation of the two parameters in the utility computation is necessary, by means of a $Balance(k)$ function that can be used to formalise these requirements:

\[
 Balance(k) =
\begin{cases}
\frac{1}{3} \cdot k + (\frac{2}{3} \cdot k \cdot DispBalance) & \textit{if } k = k_c \\
\frac{1}{3} \cdot k + (\frac{2}{3} \cdot k \cdot (1 - DispBalance)) & \textit{if } k = k_g \vee k=k_i\\
k & \textit{otherwise}
\end{cases}
\]

where $k_i$, $k_g$ and $k_c$ are the weighted parameters $U_a(c)$ and 

\[
DispBalance =  tanh \left(  \displaystyle \frac{Disp(Group)}{\delta} \right) 
\]

is another function that works on the value of group dispersion as the relationship between the area and the size of the group, applying on it the hyperbolic tangent. The value of $\delta$ is a constant that essentially represents a threshold above which the adaptation mechanism starts to become more influential; after a face validation phase, we set this value to 2.5, allowing the output of $DispBalance$ function in the range $[0,1]$ according to all elements in $U_a(c)$. The hyperbolic tangent approaches value $1$ when $Disp(Group)$ approaches $6$ (values $\geq 6$ indicate a high level of dispersion for small-medium size groups (1-4 members)).

\begin{figure}[t]
\centering
\includegraphics[width=0.6\columnwidth]{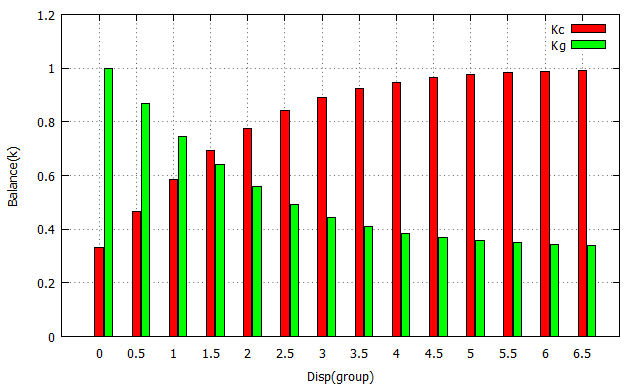}
 \caption{Graphical representation of $Balance(k)$, for $k=1$ and $\delta = 2.5$}
 \label{fig:balance}
\end{figure}

A graphical representation of the trade-off mechanism is shown in Fig. \ref{fig:balance}: red and green boxes represent the progress of parameter $k_c$ and parameter $k_g$ ($k_i$ is treated analogously), respectively. Note that the increasing of the dispersion value produces an increment of $k_c$ value and a reduction of $k_g$ parameter.

It must be emphasised the fact that this adaptive balancing mechanism and the current values for its parameters were heuristically established and they actually require a validation (and plausibly a subsequent calibration) by comparing results achieved with this configuration and relevant empirical data about group dispersion gathered from actual observations and experiments in controlled situations. 

\subsection{Simulation Results in Benchmark Scenarios}

This section describes the results of a simulation campaign carried out to evaluate the performance of the above described model that had mainly two goals: (i) \emph{validate} the model, in situations for which the adaptation mechanism was not activated (i.e. no simple groups were present in the simulated population), (ii) \emph{evaluate} the effects of the introduction of simple groups, performing a qualitative face validation of the introduced adaptation mechanism considering available video footages of the behaviour of groups in real and experimental situations. 

\begin{figure}[t]
\centering
\includegraphics[width=0.85\columnwidth] {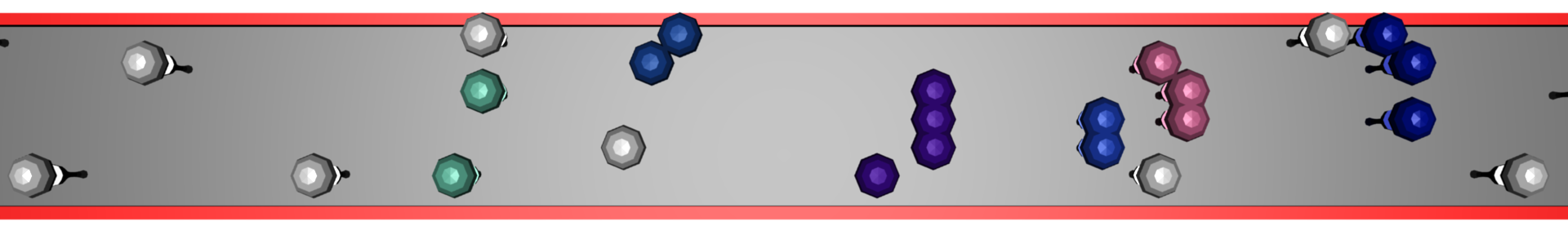}
\includegraphics[width=0.85\columnwidth] {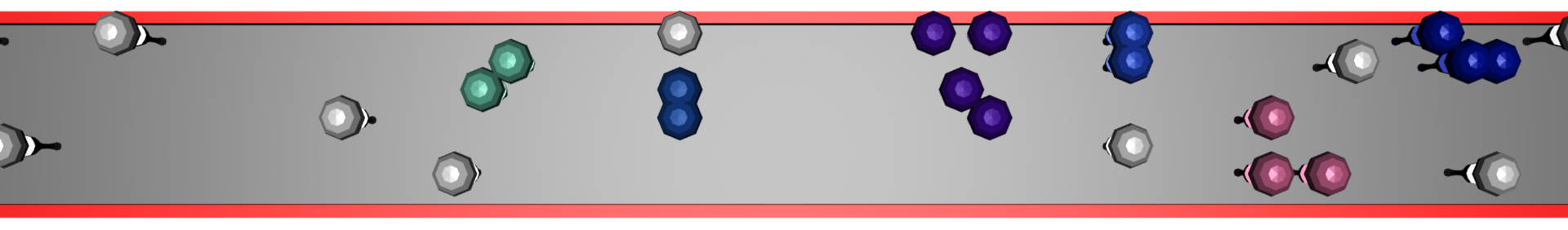}
\caption{Screenshots of the simulated corridor (A) with groups of different size in low density situations}
\label{fig:screenshot}
\end{figure}

The chosen situations are relatively simple ones and they were chosen due to the availability of relevant and significant data from the literature. In particular, we describe here a linear scenario: a \emph{corridor} in which we test the capability of the model to correctly reproduce a situation in which two groups of pedestrians enter from one of the ends and move towards the other. This situation is characterised by a counterflow causing local situations of high density and conflicts on the shared space\footnote{The term conflict used here is not associated to the issues arising from a parallel update in a CA approach: actual conflicts are prevented by the shuffled sequential update strategy. However, the choices of pedestrians are limited by the actions of the others, so this conflict is actually a conceptual one.}. We essentially evaluate and validate this scenario by means of a fundamental diagram as defined by~\cite{DBLP:reference/complexity/SchadschneiderKKKRS09}: it shows how the average velocity of pedestrians in a section (e.g. one of the ends of a corridor) varies according to the density of the observed environment. Since the flow of pedestrians is directly proportional to their velocity, this diagram is sometimes presented in an equivalent form that shows the variation of flow according to the density. In general, we expect to have a decrease in the velocity when density grows, due to the growing frequency of ``collision avoidance'' movements by the pedestrians; the flow, instead, initially grows, since it is also directly proportional to the density, until a certain threshold value is reached (also called \emph{critical density}), then it decreases. Despite being of great relevance, different experiments gathered different values of empirical data: while there is a general consensus on the shape of the function, the range of the possible values has even significant differences from different versions.

We investigated this scenario with a significant number of simulations, varying the level of density by adjusting the number of pedestrians present in the environment, so as to analyse different crowding situations. For every scenario, in terms of environmental configuration and level of density, a minimum of 3 and a maximum of 8 simulations were executed, according to the variability of the achieved results (more simulations were run when the variability was high, generally around levels of density close to the critical thresholds). Every simulation considered at least 1800 simulation turns, corresponding to 10 minutes of simulated time. The rationale was to observe a good number of complete paths of pedestrians throughout the environment, that was configured to resemble a torus (e.g. pedestrians completing a movement through it re-entered the scenario from their starting point), therefore simulations of situations characterised by a higher density were also set to last longer.

\begin{figure}[t]
\centering
\includegraphics[width=0.65\columnwidth] {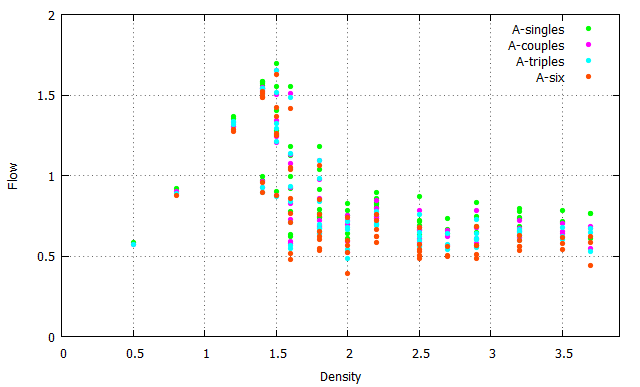}
\caption{Comparison among fundamental diagrams on aggregate data with respect to groups of different size in corridor A}
\label{fig:av_gruppi}
\end{figure}

As suggested above, we adopted two different experiment settings: in the first one, the individual pedestrians belonging to a given flow (i.e. all the pedestrians entering the corridor) are represented as members of a large structured group, but no simple groups are present. This first part of the experimentation was also necessary to perform a proper calibration of the model, for the parameters not involved in simple group modelling. In the second experimental setting, we included a variable number of simple groups (based on the total number of pedestrians in the environment and according to available data on the frequency of groups of different size in a crowd as mentioned in~\cite{TheraulazGroup} and \cite{DBLP:conf/acri/FedericiGMV12})\footnote{Groups of size 2 include about 28\% of the total number of pedestrians, groups of size 3 about 24\% and groups of size 6 about 12\%.} first of all to calibrate and qualitatively validate the adequacy of the adaptation mechanism and then to explore its implications on the overall crowd dynamics.

\begin{figure}[t]
\centering
\includegraphics[width=0.62\columnwidth] {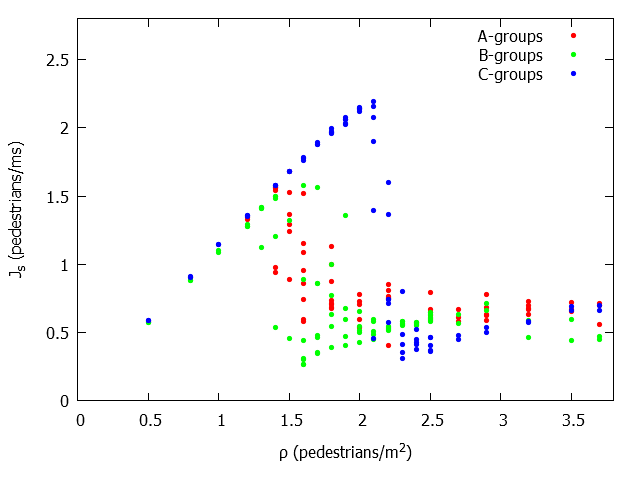}
\caption{Comparison among corridor widths: fundamental diagrams in case of groups in corridors A, B and C}
\label{fig:comparison_width}
\end{figure}


We actually simulated the linear counterflow situation in three different corridors, of growing width: their size is respectively $2.4$ m $\times$ $20$ m (A), $3.6$ m $\times$ $13.2$ m (B) and $4.8$ m $\times$ $10$ m (C). Note that the variation in terms of width and height were applied according to the choice of maintaining the total area at a level about $48$ m$^2$ in every scenario. A screenshot of the simulation scenario in low density situations including groups of different size is shown in Figure~\ref{fig:screenshot}. The choice of evaluating the influence of groups in different linear scenarios was also inspired by~\cite{zhang2011}, in which a comparison in terms of pedestrians flow from experimental data among three corridors of width $1.8$ m, $2.4$ m and $3.0$ m is presented. In this case, authors show that, in conformance with~\cite{hankin1958passenger}, above a certain minimum of about $1.22$ m, the maximum flow is directly proportional to the width of the corridors.

The model was calibrated to achieve, in situations not including simple groups, results in tune with empirical data available from the literature. We will now focus on some partly counterintuitive results in presence of simple groups. To do so, data related to the different types of simulated groups were aggregated, and a comparison among the related fundamental diagrams was performed for the movement of groups in corridor A. As summary, Fig.~\ref{fig:av_gruppi} represents on the same chart all group contributions: the depicted points represent the average flow achieved for that kind of group in the total simulated time, and generally more points are available (representing different simulation runs) for the same configuration. The overall flow of individuals is generally higher than that of groups in almost all situations, and in general with the growth of the size of a simple group we observe a decrease of its overall flow. Moreover, differences tend to decrease and almost disappear after the critical density (about $1.5$ pedestrians per square metre) is reached.

We also analysed the effect of the width of the corridor on the flow of groups (and in general on the overall pedestrian flow): Figure~\ref{fig:comparison_width} shows the different fundamental diagrams associated to all simple groups (irrespectively of their size) in corridors A, B and C. It is apparent that the critical density moves to higher levels with the growth of the corridor width, in tune with the already mentioned results discussed in~\cite{hankin1958passenger}.


\begin{figure}[t]
\centering
\includegraphics[width = 0.62\columnwidth]{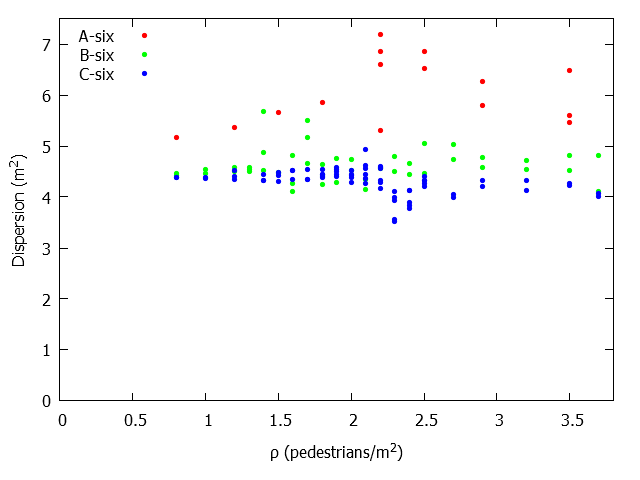}
\caption{Comparison among the level of dispersions for group of six members in corridor A, B and C}
\label{fig:dispersion_comparison}
\end{figure}

Finally, we analysed how the level of group dispersion, computed by means of the same function employed to manage the adaptive mechanism for group cohesion, varies with a changing density in the environment. The motivations of this analysis are twofold: first of all, we wanted to understand if the adaptive mechanism for group cohesion is effective, then we wanted to gather empirical data to understand if it produces 	\emph{plausible} results, in line with \emph{observed data} that, at the moment of the simulation campaign, were still not available. Figure~\ref{fig:dispersion_comparison} shown the variation of the level of dispersion for groups of size $6$ in corridors A, B and C: we can conclude that the mechanism for preserving the cohesion of simple groups is actually effective, since the growth of density does not cause a significant growth of the dispersion. On the other hand, at that moment we could not conclude that the model produces realistic results.

\section{An Analysis of Field Data About Pedestrian Crowd Dynamics}\label{sec:observation}

This Section comprises several empirical studies aimed at investigating pedestrian crowd dynamics in the natural context by using on-field observation. In particular the survey was aimed at studying the impact of grouping and proxemics behaviour on the whole crowd pedestrian dynamics. Data analyses were focused on: (i) \emph{level of density and service}, (ii) \emph{presence of groups} within the pedestrian flows, (iii) group proxemic \emph{spatial arrangement}, (iv) \emph{trajectories and walking speed} of both singles and group members. Furthermore the \emph{spatial dispersion} of group members while walking was measured in order to propose an innovative empirical contribution for a detailed description of group proxemics dynamics while walking.

The survey was performed the last 24\textsuperscript{th} of November 2012 from about 2:50 pm to 4:10 pm. It consisted in the observation of the bidirectional pedestrian flows within the Vittorio Emanuele II gallery, a popular commercial-touristic walkway situated in the Milan city centre (Italy). The gallery was chosen as a crowded urban scenario, given the large amount of people that pass through it during the weekend for shopping, entertainment and visiting touristic-historical attractions in the centre of Milan.

The team performing the observation was composed of four people. Several preliminary inspections were performed to check the topographical features of the walkway. The balcony of the gallery, that surrounds the inside volume of the architecture from about ten meters in height, was chosen as location thanks to possibility to (i) position the equipment for video footages from a quasi-zenithal point of view and (ii) to avoid as much as possible to influence the behaviour of observed subjects, thanks to a railing of the balcony partly hiding the observation equipment. The equipment consisted of two professional full HD video cameras with tripods. The existing legislation about privacy was consulted and complied in order to exceed ethical issues about the privacy of the people recorded within the pedestrian flows. 

\begin{figure}
\centering
\includegraphics[width=\textwidth]{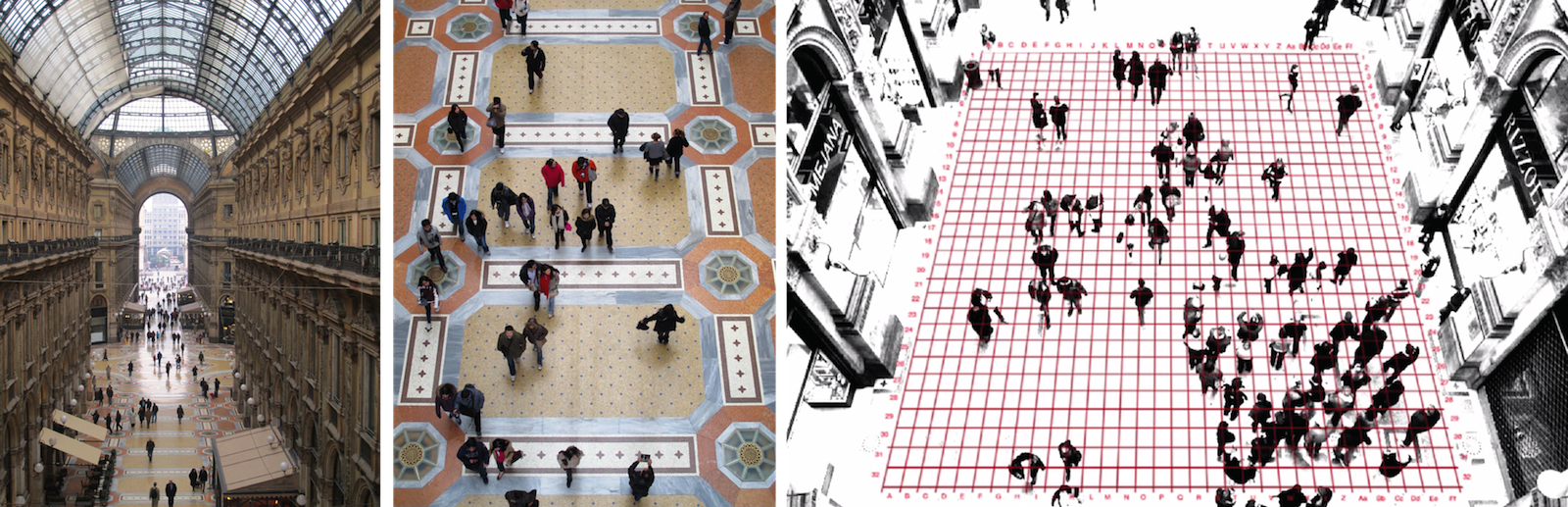}
\caption{From the left: an overview of the Vittorio Emanuele II gallery, the streaming of passerby within the walkway and a snapshot of the recorded video images with the superimposed grid for data analysis}
\label{fig:grid}
\end{figure}

Two independent coders performed a manual data analyses, in order to reduce errors by crosschecking their results. A squared portion of the walkway was considered for data analysis: 12.8 meters wide and 12.8 meters long (163.84 squared meters). In order to perform data analyses, the inner space of the selected area was discretised in cells by superimposing a grid\footnote{The grid was design by using \emph{Photoshop CS5} (according to the perspective of the video images). An alphanumeric code (from 1 to 32 on both left and right sides, and from A to Ff on both top and bottom sides) was added on the sides of the grid. Finally, the grid with a transparent background was superimposed to a black-white version of the video images by means of \emph{iMovie}. To perform counting activities, the video was reproduced by using \emph{VLC} player thanks to its possibility to: playback the images in slow motion and/or frame by frame and to use an extension time format that included hundredths of a seconds.} on the video (see Fig.~\ref{fig:grid}); the grid was composed of 1024 squares 0.4 meters wide and 0.4 meters long.

\subsection{Level of Density and Service}
The bidirectional pedestrian flows (from North to South and vice versa) were manually counted minute by minute: 7773 people passed through the selected portion of the Vittorio Emanuale II Gallery from 2:50 pm to 4:08 pm. The average level of density within the selected area (defined as the quantitative relationship between a physical area and the number of people who occupy it) was detected considering 78 snapshots of video footages, randomly selected with a time interval of one minute. The observed average level of density was low (0.22 people/squared meter). Despite it was not possible to analyse continuous situations of high density, several situation of irregular and local distribution of high density were detected within the observed scenario.

According to the Highway Capacity Manual by~\cite{milazzo_ii_quality_1999}, the level of density in motion situation was more properly estimated taking into account the bidirectional walkway level of service criteria: counting the number of people walking through a certain unit of space (meter) in a certain unit of time (minute). The average level of flow rate within the observed walkway scenario belongs to a \emph{B} level (7.78 ped/min/m) that is associated with an irregular flow in low-medium density condition.

\subsection{Flow Composition}\label{sec:flow-comp}
The second stage of data analysis was focused on the detection of groups within the pedestrian flows, the number of group members and the group proxemics spatial arrangement while walking. The identification of groups in the streaming of passerby was assessed on the basis of verbal and nonverbal communication among members: visual contact, body orientation, gesticulation and spatial cohesion among members. To more thoroughly evaluate all these indicators the coder was actually encouraged to rewind the video and take the necessary time to tell situations of simple local (in time an space) similar movements, due to the contextual situation, by different pedestrians from actual group situations. The whole video was sampled considering one minute every five: a subset of 15 minutes was extracted and 1645 pedestrians were counted (21.16\% of the total bidirectional flows). Concerning the flow composition, 15.81\% of the pedestrians arrived alone, while the 84.19\% arrived in groups: 43.65\% of groups were couples, 17.14\% triples and 23.40\% larger groups (composed of four or five members). Large structured groups, such as touristic committees, that were present in the observed situation, were analysed considering sub-groups.

\subsection{Group Proxemics Spatial Arrengement}
Results about group proxemics spatial arrangement while walking showed that:\begin{itemize}
\item 94.43\% of couples was characterised by line-abreast pattern while 5.57\% by river-like pattern;
\item 31.91\% of triples was characterised by line-abreast pattern, 9.57\% by river-like pattern and 58.51\% by V-like pattern;
\item 29.61\% of four-person groups is characterised by line-abreast pattern, 3.19\% by river-like pattern, 10.39\% by V-like pattern, 10.39\% triads followed by a single person, 6.23\% single individual followed by a triad, 7.79\% rhombus-like pattern (one person heading the group, followed by a dyad and a single person), 32.47\% of the groups split into two dyads.
\end{itemize}

\subsection{Trajectories and Walking Speed}
The walking speed of both singles and group members was measured considering the path and the time to reach the ending point of their movement in the monitored area (corresponding to the centre of the cell of the last row of the grid) from the starting point (corresponding to the centre cell of the first row of the grid). Only the time distribution related to the \emph{B} level of service was considered (as mentioned, the 59\% of the whole video footages), in order to focus on pedestrian dynamics in situation of irregular flow. A sample of 122 people was randomly extracted:  30 singles, 15 couples, 10 triples and 8 groups of four members. The estimated age of pedestrians was approximately between 15 and 70; groups with accompanied children were not taken into account for data analyses. About gender, the sample was composed of 63 males (56\% of the total) and 59 females (44\% of the total). Differences in age and gender were not considered in this study.

\begin{figure}[h!]
\centering
\includegraphics[width=\textwidth]{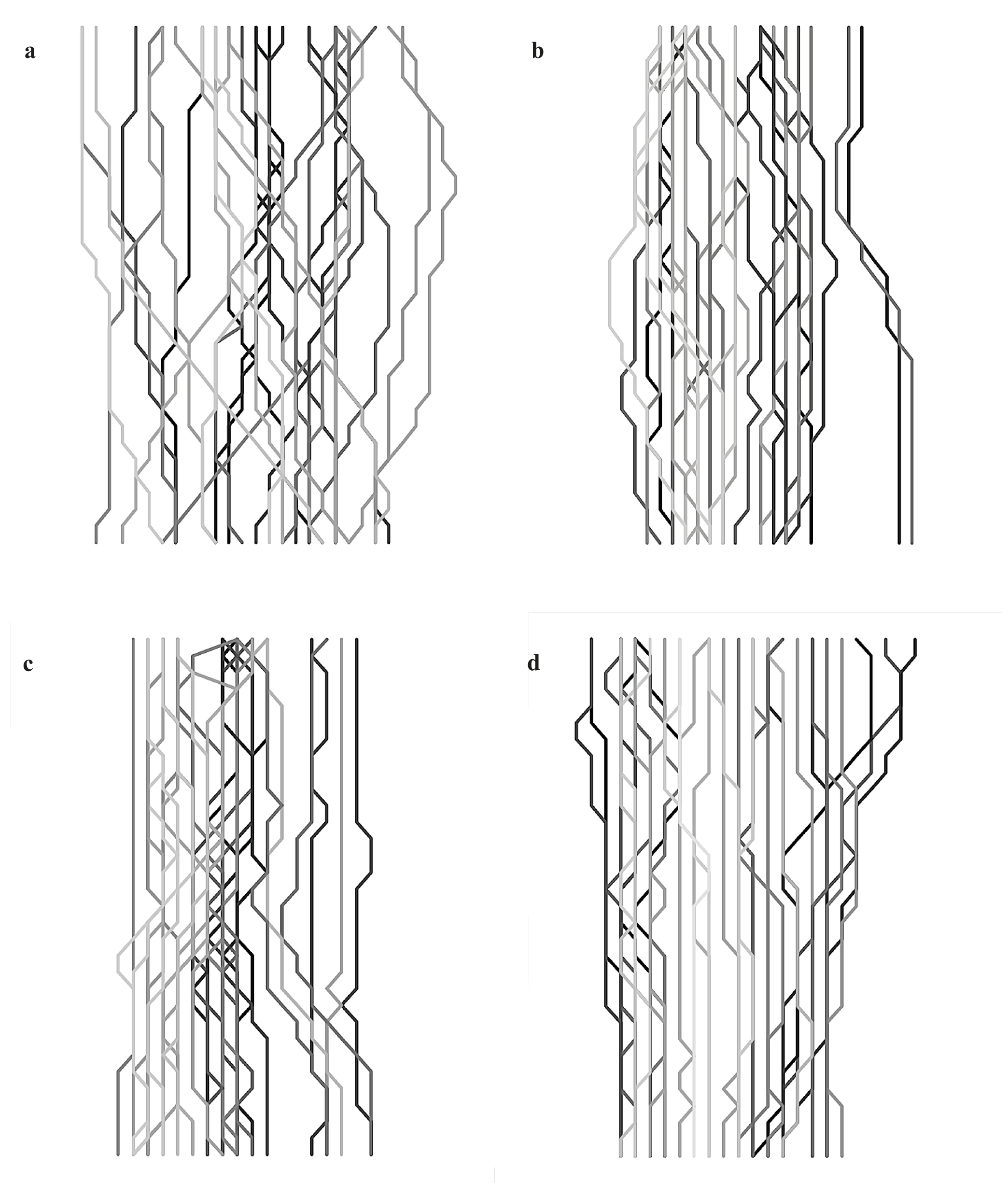}
\caption{The trajectories of the sampled pedestrian within the bidirectional flows: singles (a), couple (b), triple (c) and groups of four members (d)}
\label{fig:trajectories}
\end{figure} 

The alphanumeric grid was used to track the trajectories of both single and group members within the walkway and to measure the length of their path\footnote{To measure the walking path and speed we considered each pedestrian as a point without mass in a two-dimensional plane. By using the alphanumeric grid, we considered the cell occupied by the feet of each pedestrian as its own actual position. The starting and final steps were measured from the half of the cell, consequently 0.2 m is the corresponding length of the each related path; any diagonal step cell by cell was measured as the diagonal between the two cells (0.56 m); any straight step was measured as the segment between the centre of two cells (0.4 m).} (considering the features of the cells: 0.4 m wide, 0.4 m long) (see Fig.~\ref{fig:trajectories}).

A first analysis was devoted to the identification of the length of the average walking path of singles (M=13.96 m, $\pm$1.11), couples (M=13.39 m, $\pm$ 0.38), triples (M=13.34 m, $\pm$ 0.27) and groups of four members (M=13.16 m, $\pm$ 0.46). Then, the two tailed t-test analyses were used to identify differences in path among pedestrian. Results showed a significant difference in path length between: singles and couples (p value$<$0.05), singles and triples (p value$<$0.05), singles and groups of four members (p value$<$0.05). No significant differences were detected between path length of couples and triples (p value$>$0.05), triples and groups of four members (p value$>$0.05), couples and groups of four members (p value$>$0.05). The results showed that the path of singles is 4,48\% longer than the average path of group members (including couples, triples and groups of four members).

The walking speed of both singles and group members was detected considering the path of each pedestrian within the flows and the time to reach the ending point from their starting point. A first analysis was devoted to the identification of the average walking speed of singles (M=1.22 m/s, $\pm$ 1.16), couples (M=0.92 m/s, $\pm$ 0.18), triples (M=0.73 m/s, $\pm$ 0.10) and groups of four members (M=0.65 m/s, $\pm$ 0.04). Then, the two tailed t-test analyses were used to identify differences in walking speed among pedestrian. Results showed a significant difference in walking speed between: singles and couples (p value$<$0.01), singles and triples (p value$<$0.01), singles and groups of four members (p value$<$0.01), couples and triples (p value$<$0.01), triples and groups of four members (p value$<$ 0.05). In conclusion, the results showed that the average walking speed of group members (including couples, triples and groups of four members) is 37.21\% lower than the walking speed of singles.

The correlated results about pedestrian path and speed showed that in situation of irregular flow singles tend to cross the space with more frequent changes of direction in order to maintain their velocity, avoiding perceived obstacles like slower pedestrians or groups. On the contrary, groups tend to have a more stable overall behaviour, adjusting their spatial arrangement and speed to face the contextual conditions of irregular flow: this is probably due to (i) the difficulty in coordinating an overall change of direction and (ii) the tendency to preserve the possibility of maintaining cohesion and communication among members.  

\subsection{Group Proxemics Dispersion}
In order to improve the understanding of pedestrian proxemics behaviour the last part of the study is focused on the dynamic spatial dispersion of group members while walking. The dispersion among group members was measured as the summation of the distances between each pedestrian and the centroid (the geometrical centre of the group) all normalised by the cardinality of the group. The centroid was obtained as the arithmetic mean of all spatial positions of the group members, considering the alphanumeric grid. In order to find the spatial positions, the trajectories of the group members belonging to the previous described sample (15 couples, 10 triples and 8 groups of four members) were further analysed. In particular, the positions of the group members were detected analysing the recorded video images every 40 frames (the time interval between two frames corresponds to about 1.79 seconds, according to the quality and definition of the video images) starting from the co-presence of the all members on the alphanumeric grid. This kind of sampling permitted to consider 10 snapshots for each groups.

A first analysis was devoted to the identification of the average proxemics dispersion of couples (M=0.35 m, $\pm$ 0.14), triples (M=0.53 m, $\pm$ 0.17) and groups of four members (M=0.67 m, $\pm$ 0.12). Then, the two tailed t-test analyses were used to identify differences in proxemics dispersion among couples, triples and groups of four members. Results showed a significant difference in spatial dispersion between: couples and triples (p value$<$0.05), couples and groups of four members (p value$<$0.01). No significant differences between triples and groups of four members (p value$>$ 0.05). In conclusion, the results showed that the average spatial dispersion of triples and groups of four members while walking is 40.97\% higher than the dispersion of couples. 

In order to be able to provide useful indications for the calibration of the adaptive simple group cohesion mechanism we also evaluated the average dispersion of the observed groups in terms of area covered by the group: due to the discretisation of the pedestrian localisation mechanism, we were able to essentially count the occupied cells by a sort of convex hull computed considering the positions of pedestrians as vertexes, analogously as for the dispersion metric defined and employed in the simulation model. The results of this operation estimated the dispersion of couples (Ma=0.6 m$^2$), triples (Ma=0.8 m$^2$) and groups of four (Ma=1.3 m$^2$). These values are currently being employed to calibrate and validate the simple group cohesion mechanism in the conditions of relatively low density and irregular flow. 

\begin{figure}[h!]
\centering
\includegraphics[width=.65\textwidth]{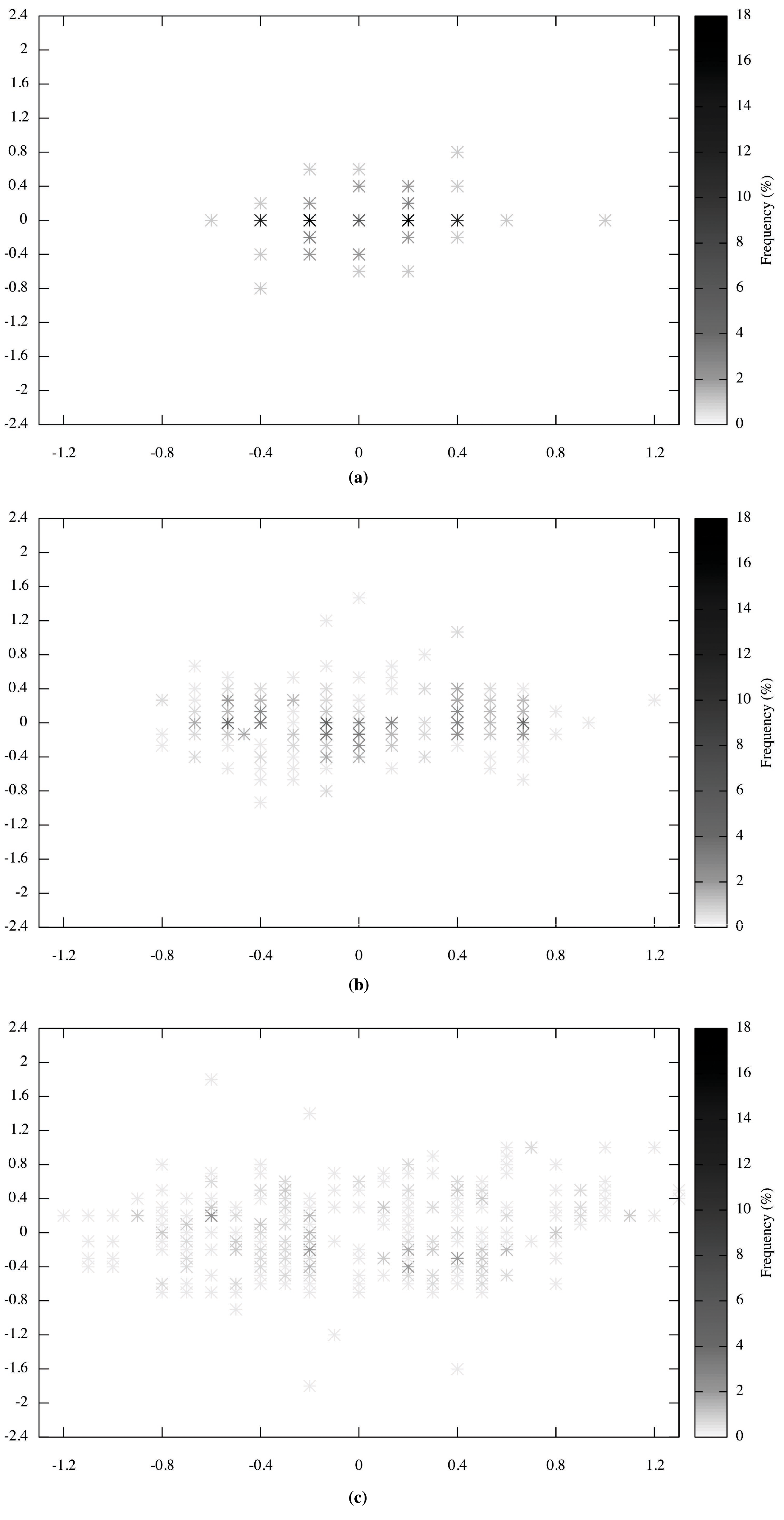}
\caption{A diagram showing most frequent positions, normalised with respect to the centroid and the movement direction, assumed by members of couples (a), triples (b) and groups of four members (c).}
\label{fig:scatter-all}
\end{figure}

Starting from the achieved results about group proxemics dispersion, we finally focused on a quantitative and detailed description of group spatial layout while walking. The normalised positions of each pedestrian with respect to the centroid and the movement direction were detected by means of a sample of 10 snapshots for each groups (15 couples, 10 triple and 8 groups of four members) and then further analysed in order to identify the most frequent group proxemics spatial configurations, taking into account the degree of alignment of each pedestrian (see Figure~\ref{fig:scatter-all}). Result showed that couple members tend to walk side by side, aligned to the each other with a distance of 0.4 m (36\% of the sample) or 0.8 m (24\% of the sample), forming a line perpendicular to the walking direction (line abreast pattern); triples tend to walk with a line abreast layout (13\% of the sample), with the members spaced of 0.60 m. Regarding groups of four members it was not possible to detect a typical spatial pattern: the reciprocal positions of group members appeared much more dispersed than in the case of smaller groups, probably to due the continuous arrangements in spatial positioning while walking. The results are in line with the previously described spatial arrangements related to the total observed pedestrian flows (see Section~\ref{sec:flow-comp}), representing an innovative contribution for the understanding of group proxemics dynamics in motion situation, once again in situations characterised by a relatively low density and irregular flow.

Empirical data about high density situations would be necessary to actually tune the mechanism in moderate and high density situations, but in this kind of scenario this observation and analysis framework would very likely be inappropriate and ineffective. A more sophisticated and at least partly automated (employing computer vision techniques and maybe machine learning approaches to support the detection of groups) controlled experiments (given the high density situation, some help to the tracking and detection approaches would probably be necessary in terms, for instance, of markers to highlight the heads of group members) is probably needed to actually face the challenges of acquiring empirical evidences about the behaviour of groups in high density situations. The participation of psychologists in the definition of such an experimental observation setting would also help in managing any kind of tiring and learning effect due to the experimental procedure.  

\section{Conclusion and Future Developments}

The paper discussed an integrated approach to the analysis and synthesis of pedestrian and crowd behaviour, in which the two aspects are actually set in an integrated framework and they mutually benefit in different ways. A general schema describing the conceptual pathways connecting modelling and analysis steps was described. A case study in which modelling and simulation approaches, specifically focused on adaptive group behaviour enabling the cohesion of simple groups, produced a research question and some usable techniques to crowd analysis approaches was also introduced. On the other hand, a subsequent observation and analysis about the phenomenology represented in the model was also described. Currently the empirical evidences resulting from this analysis activity are being used to validate and calibrate the model for group cohesion. In particular the simulation model correctly reproduces some of the observed phenomena, in particular, lower walking speeds for groups and tendency to preserve cohesion (although this aspect is undergoing a further calibration). Additional elements that are now being evaluated are related to the capability of the model to generate spatial patterns resulting from the analysis.

Moreover, the analysis of the gathered video footage also highlighted additional phenomenologies that are now being more thoroughly analysed: in particular, patterns of ``leader--follower'' behaviour within groups were detected and the introduced simulation model presents all the necessary elements and mechanisms to represent this kind of pattern. Future works are also aimed at supporting innovative automated analysis techniques based on computer vision and machine learning approaches.

\section*{Acknowledgements}
The authors would like to thank Milano's Municipality for granting the necessary authorisations to carry out the observation in the Vittorio Emanuele II Gallery. 


\end{document}